\documentclass[nonacm,acmsmall]{acmart}

\usepackage{longtable} 
\usepackage{pdflscape} 
\usepackage{hyperref}
\usepackage{cleveref}

\AtBeginDocument{%
  \providecommand\BibTeX{{%
    \normalfont B\kern-0.5em{\scshape i\kern-0.25em b}\kern-0.8em\TeX}}}

\begin{document}

\title{Using Spectral Graph Theory to Map Qubits onto Connectivity-Limited Devices}

\author{Joseph X. Lin}
\email{joelin@mit.edu}
\author{Eric R. Anschuetz}
\orcid{0000-0002-9825-3692}
\email{eans@mit.edu}
\author{Aram W. Harrow}
\orcid{0000-0003-3220-7682}
\email{aram@mit.edu}
\affiliation{%
  \institution{Massachusetts Institute of Technology}
  \streetaddress{77 Massachusetts Avenue}
  \city{Cambridge}
  \state{Massachusetts}
  \postcode{02139}
}

\renewcommand{\shortauthors}{Lin, Anschuetz, and Harrow}

\begin{abstract}
  We propose an efficient heuristic for mapping the logical qubits of quantum algorithms to the physical qubits of connectivity-limited devices, adding a minimal number of connectivity-compliant SWAP gates. In particular, given a quantum circuit, we construct an undirected graph with edge weights a function of the two-qubit gates of the quantum circuit. Taking inspiration from spectral graph drawing, we use an eigenvector of the graph Laplacian to place logical qubits at coordinate locations. These placements are then mapped to physical qubits for a given connectivity. We primarily focus on one-dimensional connectivities, and sketch how the general principles of our heuristic can be extended for use in more general connectivities.
\end{abstract}



\maketitle

\section{Introduction}~\label{chap:one}
The field of quantum computation has shown immense promise for solving certain problems more efficiently than classical computers including prime factorization~\cite{shor1999polynomial}, unstructured search~\cite{grover1997quantum}, optimization~\cite{farhi2014quantum}, and chemical simulation~\cite{olson2017quantum}. However, while advantageous quantum algorithms have been thoroughly developed in theory, the technological implementations of quantum devices is still very much in its infancy. The challenges and resource constraints of current and near-term devices present roadblocks for the feasibility of practial quantum information processing.

Because quantum entangling operations are crucial to setting quantum computing apart from classical computing, it is important in many practical algorithms for operations to be applied between multiple qubits. However, on many present day quantum architectures, the application of multiqubit gates is not always possible among every subset of qubits. This particular challenge has to do with the \emph{connectivity limitation} of physical devices, which constrains the set of allowable operations. Therefore, there is a compilation step needed to convert a theoretical quantum algorithm, which often assumes full connectivity between qubits, to an equivalent \emph{connectivity compliant} circuit that can be run on a physical device. In this work, we focus on this challenge and develop a novel approach using spectral graph theory.

In Sec.~\ref{chap:two}, we provide the background and motivation for our problem. This includes formally describing the problem of converting quantum circuits to be compliant with the connectivity constraints of a device, surveying prior solutions to the problem, and reviewing relevant results from spectral graph theory. Next, we explain our algorithm in Sec.~\ref{chap:three}, discussing design principles that motivate our decisions, specific details for practical implementation, and overall runtime. In Sec.~\ref{chap:four} we discuss the use of known benchmarks and a comparable open source algorithm to evaluate our own algorithm, and show the results in Sec.~\ref{chap:five}. Finally, we conclude in Sec.~\ref{chap:six} and provide future directions for improvement.

\section{Background}\label{chap:two}
In this section, we provide the necessary background for our algorithm. We first formalize the problem of making a quantum circuit connectivity-compliant for a given device. Then, we survey related work on the problem, and close with a description of spectral graph theory and token swapping, two key components to our algorithm.

\subsection{Problem Description}\label{sec:problem-desc}
As mentioned in Sec.~\ref{chap:one}, our objective is to start with a circuit containing only single qubit and CNOT gates, with no connectivity constraints, and transform the circuit into one adhering to the connectivity constraints of the device. To differentiate the two domains, we call the untransformed circuit the \emph{logical} circuit and the transformed circuit the \emph{physical} circuit. We say that the logical circuit operates on logical qubits with logical gates, and similarly for the physical circuit. 

For this transformation to be meaningful, the circuits must be equivalent in the sense that their unitary descriptions are equal. There are certain operations that can be performed that lead to equivalent circuits in all cases.  These include:
\begin{enumerate}
	\item the commutation of disjoint gates. When two gates operate on disjoint subsets of qubits, the order in which they are applied does not matter. 
	\item the logical reordering of qubits. Circuits are equivalent under the relabling of qubits.
	\item the physical reordering of qubits. When a SWAP gate is inserted for qubits $i$ and $j$, and gates and measurements are changed to take the SWAP into account, the circuit is equivalent.
        \end{enumerate}
        See \cite{NRSCM} for a longer and more thorough list of circuit equivalences. 
We assume that the first two operations can be done with no cost, but that the addition of SWAP gates should be minimized to save resources and to minimize the effects of noise in physical implementations of the circuit. 

We are now ready to formalize the connectivity-compliance problem. As previously discussed, only multi-qubit gates of the circuit are affected by connectivity; therefore, without loss of generality we consider only the CNOT gates of a circuit. Let $Q$ be the set of logical qubits of the circuit, with $M\equiv\left\lvert Q\right\rvert$. Similarly, let the set $C$ be the CNOTs of a logical circuit, with $N\equiv\left\lvert C\right\rvert$. We can represent $C$ as an ordered list of pairs $\left[\left(l_1^c, l_1^t\right), \ldots, \left(l_N^c, l_N^t\right)\right]$, where $l_i^c\in Q$ and $l_i^t\in Q$ are the logical control and target qubits, respectively, of the $i$-th CNOT.

Next, we represent the connectivity characteristics of an architecture by a graph $G = \left(V, E\right)$, where the vertices are the physical qubits and the edges represent pairs of qubits which support two qubit operations. We assume $\left\lvert V\right\rvert = M$. We also assume that the graph is unweighted and undirected, meaning each edge can support a CNOT in either direction and can do so with equal ``ability'' (e.g. fidelity, time, etc.) as any other edge.

Now, let $C^{\prime}$ be the ordered list of two-qubit gates of another circuit, where $N^{\prime}=\left\lvert C^{\prime}\right\rvert$. It also can be represented as an ordered list of tuples $\left[\left(p_1^1, p_1^2, \textrm{type}_1\right), \ldots, \left(p_{N^{\prime}}^1, p_{N^{\prime}}^2, \textrm{type}_{N^{\prime}}\right)\right]$, where $\textrm{type}_i$ represents the type of gate that gate $i$ is (i.e. either CNOT or SWAP), and $p_i^1, p_i^2\in V$ represent the two physical qubits on which the gate is applied.

If we can transform circuit $C$ to $C^{\prime}$ using the three equivalency rules established above, then the circuits are equivalent. Furthermore, if the physical qubits on each gate of $C^{\prime}$ are connected, i.e. $\left(p_i^1, p_i^2\right) \in E$ for all $i$, then the transformed circuit is connectivity compliant according to $G$. We will deem a transformation \emph{valid} if an only if both criteria are satisfied: the transformed circuit must be both equivalent to the original circuit and also compliant to the connectivity rules.

Finally, our objective is to find a valid $C^{\prime}$ that minimizes $N^{\prime}$; as $C$ and $C^{\prime}$ must have the same number of CNOTs assuming $C^{\prime}$ was constructed using the aforementioned circuit transformation rules then, by construction, this optimization minimizes the total number of added SWAP gates.

A couple notes should be made about the $C^{\prime}$ we consider. First, no transformed circuit should have a SWAP gate that can be commuted to the front of the circuit. This is because another equivalent and still valid circuit can be made by replacing that SWAP with a logical relabeling (that is, using transformation rule 2 instead of 3). Similarly, no SWAP gates should be able to be commuted to the end of the circuit. This is because these SWAP gates can be removed and any single qubit gate or measurements happening at the end of the circuit can be logically reassigned.

To finish our problem description, we discuss two different ways to view the transformation problem. The first is SWAP based, and is essentially how we described this problem: add SWAP gates to $C$ (and adjust any affected gates that follow) until the resulting $C^{\prime}$ is connectivity-compliant. Implicit in this was also deciding on an initial one-to-one logical relabeling $\pi_0 : Q\rightarrow V$, which is the mapping at the beginning of $C^{\prime}$. 

The second interpretation is layer and permutation based. Note that $C^{\prime}$ can be viewed as layers of CNOT gates which can be applied using various $Q\rightarrow V$ permutations, with the intermediate SWAP gates acting as a bridge between permutation. This can be seen as follows. Suppose we took $C^{\prime}$, and commuted the gates so that as many CNOTs were at the front as possible before a SWAP gate must be applied. Then, those CNOTs are the CNOTs from $C^{\prime}$ that can be made connectivity compliant through the initial mapping $\pi_0$. Then, suppose that we remove those CNOTs from the circuit (as they have been now applied already), and perform a similar commutation to bring as many SWAPs to the front as possible. These SWAPs form a physical permutation $\sigma_1: V\rightarrow V$. If we then remove the SWAPs and again bring as many CNOTs to the front as we can, we have another layer of CNOTs that can be applied from the logical circuit, under the permutation $\pi_1 = \sigma_1\pi_0 : Q\rightarrow V$. This can be continued until all gates of $C^{\prime}$ are exhausted.

Using this second interpretation, we can describe an equivalent way of transforming $C$ into a valid $C^{\prime}$:
\begin{enumerate}
	\item Beginning with $C$, allow commutation of CNOTs that are nonoverlapping.
	\item Partition the CNOTs into layers such that the CNOTs of each layer can be made compliant using some mapping $\pi_i: Q\rightarrow V$. 
	\item Using only compliant SWAP gates, create a physical permutation $\sigma_{i+1} V\rightarrow V$ that transfers one logical permutation to the next one, via $\pi_{i+1} = \sigma_{i+1}\pi_i$.
\end{enumerate}
The optimization problem is still the same, find the $C^{\prime}$ with the fewest number of added SWAPs, but the optimizaiton is now over  layers and permutations, rather than individual SWAPs. As we will see, this interpretation can allow for the overall optimization problem to be subdivided into smaller problems with well-known solvers.

\subsection{Prior Work}
In the literature, results and algorithms related to this issue of connectivity-compliance have gone by many names, including circuit layout, circuit transformation, qubit allocation, qubit mapping, and qubit routing~\cite{siraichi2018qubit,wille2014optimal,murali2019noise,zulehner2018efficient,childs2019circuit,li2019tackling,pedram2016layout,chakrabarti2011linear,shafaei2013optimization,finigan2018qubit}. We provide a survey of these results, and go more in-depth with a few of particular interest.

First, we establish the various choices that each algorithm designer has had to make:
\begin{enumerate}
	\item \emph{Metric to optimize}. Examples include \textbf{circuit size} (i.e. number of gates in the final circuit; often equivalent to number of added SWAP gates); \textbf{circuit depth} (number of layers in the circuit, where each qubit is acted on by only one gate per layer); and \textbf{error rate}.
	\item \emph{Exact vs. Approximate}. Designers must make the choice between a \textbf{brute force} approach versus using a \textbf{heuristic} or \textbf{relaxation}.
	\item \emph{Connectivity Constraint}. Examples include \textbf{LNN}, \textbf{2D grid} nearest neighbor, connectivities of \emph{actual NISQ devices}; \textbf{ring}; and \textbf{arbitrary} connectivity
	\item \emph{Problem Interpretation}. As described above, two common interpretations are \textbf{SWAP-based} and \textbf{layer- and permutation-based}.
	\item \emph{Solving strategy}. Many times, this involves reduction in part to another well-known problem. This includes \textbf{dynamic programming}; \textbf{search} (e.g. breadth-first and A* search); \textbf{Boolean or satisfiability solvers}; and \textbf{MinLA solvers}.
\end{enumerate}

For example, Siraichi et al.~\cite{siraichi2018qubit} provide both an exact and approximate solver to minimize the circuit size for arbitrary connectivity. The former makes use of dynamic programming, while the latter uses a search-like procedure. Both use an interpretation close to the SWAP-based one discussed in Sec.~\ref{sec:problem-desc}. Another example of an exact solver is~\cite{wille2014optimal}, which specifically solves the problem for linear nearest-neighbor architectures using SAT solvers and pseudo-Boolean optimization.

Because solving the qubit mapping problem exactly is NP-complete~\cite{maslov2008quantum}, all exact solvers have prohibitively large asymptotic runtimes, and in practice cannot be used beyond very small circuits. Therefore, we shift our focus to approximate approaches.

In 2018, IBM held a challenge~\footnote{\url{https://www.ibm.com/blogs/research/2018/08/winners-qiskit-developer-challenge/}} for compiling circuits to various architectures. The winning algorithm, by Zulehner, Paler, and Wille~\cite{zulehner2018efficient}, makes novel use of A* search to construct the SWAP gates between permutations. First, the CNOTs are divided up into layers in a greedy fashion (i.e. putting each CNOT in the left-most layer). An initial permutation is then needed; it is proposed that, instead of a random initial mapping, an empty partial mapping is used. Then when searching for a permutation for a given layer, the cost of assigning a previously unassigned physical qubit can be made $0$. The search is then from the previous permutation to any one that makes all CNOTs of the next layer connectivity compliant.

One of the second place winners expanded upon their algorithm, and proposed both a framework and several solvers~\cite{childs2019circuit} for this problem. They formalize the layer- and permutation-based interpretation of the problem, defining the concepts of \textit{permuters} and \textit{mappers}. The former is a subroutine for finding an (approximately) optimal sequence of SWAPs to go from one permutation to the next, while the latter is used to determine what each permutation is. The overall algorithm then involves invoking the mapper to generate a permutation, applying all first-layer CNOTs compliant for that permutation, and repeating the process until the unapplied CNOTs run out. Then, the permutations are bridged by invoking the permuter. For their circuit size optimizing permuter, they use a modified, approximate token swapping algorithm (we describe the original algorithm in Sec.~\ref{sec:token}). For the circuit size optimizing mappers, they present four different variations. Most consider all possible gates in the first layer and for each one finds the permutation that requires the fewest number of SWAPs, according to the permuter, while allowing that gate to be compliant. Which gate is chosen (and therefore which permutation) depends on the mapper.

SABRE~\cite{li2019tackling} is another approximate algorithm that minimizes circuit size. It is another heuristic, search-based algorithm targeting arbitrary connectivities. We mention them briefly to highlight their focus on the SWAP-based interpretation, as well as their use of look-ahead and bidirectionality. Look-ahead is the notion of using not just the first layer CNOTs but also some later layer CNOTs, with a parametric weighting to lower its importance relative to the first layer. Bidirectionality makes use of the fact that a connectivity compliant transformation of the reverse circuit, when itself reversed, becomes a connectivity compliant transformation of the original circuit reversed. Therefore, considering both directions can be useful for finding an optimal transformation.

The final algorithms we review share the use of a \emph{interaction (or adjacency) graph}. An interaction graph is a weighted, undirected graph where the vertices represent logical qubits. The graph is meant to indicate, in a sense, which qubits should be placed adjacent to one another, prioritizing higher edge weights. One approach, used by both~\cite{shafaei2013optimization,pedram2016layout} and~\cite{chakrabarti2011linear}, is as follows. Let there be an edge of weight $w_{ij}$ between logical qubits $q_i$ and $q_j$ if and only if there are exactly $w_{ij}$ two-qubit gates (i.e. CNOTs) between $q_i$ and $q_j$ in the circuit. Both sets of papers use this interaction graph to map the logical qubits onto an LNN architecture. While the former set of two papers use the interaction graph and solves a Minimum Linear Arrangement (MinLA) problem, the latter paper performs a graph partitioning algorithm on the interaction graph.

In future Sections, we will refer back to this survey to compare and contrast the design, implementation, and performance of our own algorithm.

\subsection{Theory Background}
To close this Section, we will describe the necessary theoretical background on which we will base our own algorithm. We describe the use of spectral graph theory for assigning points of a graph to coordinate locations, as well as the token swapping problem for transitioning between permutations via swapping. These have the significance of providing ways to generate the logical permutations $\pi_i: Q\rightarrow V$ and physical permutations $\sigma_i: V\rightarrow V$, respectively, that were described in Sec.~\ref{sec:problem-desc}.

\subsubsection{Spectral Graph Theory and Drawing}\label{sec:spectral}
Here, we discuss a method for mapping the vertices of a \emph{weighted} graph to Cartesian coordinate locations, given that the edge weights are some sort of priority for how close the vertices should be. We develop this method and make the description more precise below.

Let $\Tilde{G} = \left(\Tilde{V}, \Tilde{E}\right)$ be a weighted, undirected graph with vertices $\Tilde{v}_i$ for $i = 1, \ldots, n$. Next, suppose that a nonnegative weight $w_{ij} = w_{ji}$ is associated between each pair of vertices $\Tilde{v}_i$ and $\Tilde{v}_j$, where $i\neq j$. If $\left(\Tilde{v}_i, \Tilde{v}_j\right)\in \Tilde{E}$ is an edge, then $w_{ij} > 0$; otherwise, $w_{ij} = 0$. The \emph{Laplacian} of $\Tilde{G}$ can then be defined as follows:
\begin{definition}{(\textbf{Graph Laplacian})}
	The Laplacian of a graph $\Tilde{G} = \left(\Tilde{V}, \Tilde{E}\right)$ is defined by a symmetric, $n\times n$ matrix $L$ such that:
	\begin{align}
		L_{ij} = \begin{cases}
			\displaystyle\sum_{k\neq i}{w_{ik}}, & \textrm{if }i = j\\
			-w_{ij}, & \textrm{if }i\neq j
		\end{cases}
	\end{align}
\end{definition}
Note that $L$ can also be written as $D - A$, where $D$ and $A$ are the degree and adjacency matrices, respectively, of $\Tilde{G}$.

Citing~\cite{koren2003spectral}, there are several properties of the Laplacian that make it appealing for our purposes. In particular, its eigenvectors help provide a drawing for the graph that places the vertices at spatial locations while optimizing over a quantity related to the edge weights. First, consider the following result:
\begin{lemma}\label{lemma:lapl-quad}
	Let $x\in \mathbb{R}^n$. Then
	\begin{align}
		x^T L x = \sum_{i < j} w_{ij} \left(x_i - x_j\right)^2.
	\end{align}
      \end{lemma}

      Since this is a standard fact, we postpone the proof to \Cref{app:spectral-proofs}.

Next, it can be directly observed that $L$ is a real, symmetric matrix. It can further be easily shown that $L$ is positive semidefinite. This allows us to conclude that $L$ has nonnegative, real eigenvalues and real, orthogonal eigenvectors.  The lowest eigenvalue is always 0, regardless of the graph.

\begin{lemma}\label{lemma:zero-eig}
  Let $1_{n} = (1, \ldots, 1)^T\in \mathbb{R}^n$ be the all ones vector. Then:
  \begin{align}
    L1_n = 0,
  \end{align}
  i.e. $1_n$ is an eigenvector with eigenvalue 0.
\end{lemma}
Again the proof of this well-known property is in \Cref{app:spectral-proofs}.

Finally, we are ready to motivate an optimization problem relating $L$ to drawing a graph. Suppose the edge weight $w_{ij}$ is a measure of ``how important'' it is for vertices $v_i$ and $v_j$ to be near one another, with a larger weight correlating with greater importance. Next, suppose we seek to layout the vertices in 1D according to a vector $x\in \mathbb{R}^n$, with $v_i$ at location $x_i$. A reasonable minimization problem would then be:
\begin{align}
	\min_x \quad & \sum_{i < j} w_{ij} (x_i - x_j)^2\\
	\text{s.t.} \quad & \text{Var(x)} = \frac{1}{n}.
\end{align}
Note that the constraint is simply chosen to normalize the scale of the drawing; the choice of $1/n$ is completely arbitrary but helps simplify the calculations in the following steps. Next, notice that we can use Lemma~\ref{lemma:lapl-quad} to rewrite the problem as:
\begin{align}
	\min_x \quad & x^T L x \\
	\text{s.t.} \quad & \text{Var(x)} = 1/n.
\end{align}
We then invoke Lemma~\ref{lemma:zero-eig} to note that both the objective and variance do not change under a translational shift; that is, both $x$ and $x + \alpha 1_n$ for any $\alpha$ have the same variance and same objective value. Therefore, without loss of generality we choose for $x^t 1_n = 0$, i.e. that the average position is $0$. This is, again, an arbitrary choice made out of convenience, as it allows us to write:
\begin{align}
	\text{Var}(x) = x^T x / n.
\end{align}
Combining all of these results, we can write our desired optimization problem as:
\begin{align}
	\min_x \quad & x^T L x \\
	\text{s.t.} \quad & x^T x = 1\\
	& x^T 1_n = 0.
\end{align}

Define the eigenvalues of $L$ to be $0 = \lambda_1 \leq \lambda_2\leq \cdots \leq \lambda_n$ with corresponding normalized eigenvectors $1_n/\sqrt{n} = y^{\left(1\right)}, y^{\left(2\right)}, \ldots, y^{\left(n\right)}$. As our optimization problem is to minimize the Rayleigh quotient $R\left(L, x\right)=x^T L x$, subject to $x$ being normalized and over the subspace orthogonal to the eigenvector $y^{\left(1\right)}$, we can invoke the Courant--Fischer principle and immediately write down that $y^{\left(2\right)}$ is an optimal solution with optimal objective value $\lambda_2$. This special vector is known as the \emph{Fiedler vector} of $G$.

To conclude this section, we note a couple extensions of this result. First, suppose we only allowed for the vertices to be placed at discrete locations, e.g. integer locations. Formally, suppose we required a one-to-one mapping $\pi : \Tilde{V}\rightarrow \left\{1, 2, \ldots, n\right\}$. A natural way to get an approximate solution is to take the components of $y^{(2)}$ and use the order they impose:
\begin{align*}
	\pi\left(\Tilde{v}_i\right) > \pi\left(\Tilde{v}_j\right) \text{ only if } y^{\left(2\right)}_i \geq y^{\left(2\right)}_j.
\end{align*} 
We cite~\cite{juvan1992optimal,petit2003experiments,i1998approximation} for this result, which show that this mappings gives a good approximation to the related Minimum Linear Arrangement (MinLA) problem. Note that in the MinLA problem, the $\left(x_i - x_j\right)^2$ terms are replaced by $\left\lvert \varphi\left(v_i\right) - \varphi\left(v_j\right)\right\rvert$ terms. We cite this result, not because we wish to solve the MinLA problem, but because, as we previously mentioned, MinLA solvers have been used in part for approximate solvers. Therefore, we are motivated to use a solver as an alternative but related heuristic.

Finally, although we will mostly focus on using the 1D result, we note that this spectral drawing method can be expanded to two (or even more) dimensions. To do this, we again cite~\cite{koren2003spectral}, which motivates the use of $y^{\left(3\right)}$ for the second dimension. This vector solves the same optimization problem, but with the added constraint that $x$ is orthogonal to $y^{\left(2\right)}$ as well. This provides a drawing in which the two dimensions are uncorrelated, which allows for the added dimension to provide as much new information as possible. While this is useful for drawing purposes, the use of this method to map to discrete grid locations in two dimensions is less trivial than the one-dimensional case.

\subsubsection{Token Swapping}\label{sec:token}
We now discuss the subroutine that will be used to transfer logical qubits from one permutation to another using only connectivity compliant SWAP gates. Our problem is exactly equivalent to the \emph{Token Swapping Problem}, which can be described as follows.

Suppose we have a graph $G = \left(V,E\right)$ of $n$ vertices. Further suppose that we have tokens $t_1, \ldots, t_n$ that are to be placed on the vertices so that each vertex has exactly one token. Given an initial and desired final mapping of tokens to graph vertices, the problem is then to transform the former to the latter only by swapping the tokens on a pair of vertices connected by an edge.

This is exactly analogous to the connectivity compliance problem for quantum circuits: the vertices $V$ represent physical qubits, the edges $E$ are between physical qubits which allow two-qubit operations, and the tokens are the logical qubits to be mapped on the physical devices. Swapping adjacent tokens then amounts to applying SWAP gates between adjacent qubits. 

This problem is NP-hard, and the best known exact algorithm requires an exponential runtime. Therefore, we use an approximate algorithm described in~\cite{miltzow2016approximation} that gets a sequence of swaps of length within 4 times the optimal length for general graphs, and within 2 times the optimal length for trees. For completeness, we review the algorithm in Appendix~\ref{app:token-swapping}.

\section{The Spectral Mapping Algorithm}\label{chap:three}

\subsection{Design Principles and Motivations}\label{sec:design}
The related work and subroutines outlined in Sec.~\ref{chap:two} motivate the design our algorithm. We start by specifying the five general categories we previously listed as choices for algorithm designers.

First, we will be focusing on an approximate solution. As previously discussed, an exact solution takes exponentially long to solve for (and often requires exponential space as well); this is not feasible for circuits consisting of more than just a few qubits and CNOTs. Next, we adhere to the permutation-based interpretation of the problem, specifically the mapper-permuter model from~\cite{childs2019circuit}. This provides the problem with added structure and modularity, which in turn allows us to design and evaluate smaller subproblems independently. Like~\cite{childs2019circuit}, we will use the approximate token swapping algorithm as our permuter, which we described in Sec.~\ref{sec:token}.

Next, we consider the choice of mapper. Note that here we only describe the high-level design, omitting details that will be presented in later sections. We are motivated by~\cite{shafaei2013optimization,pedram2016layout,chakrabarti2011linear} in the creation of a weighted interaction graph. However, where previous uses of interaction graphs do not account for how ``far'' in the future a CNOT occurrence happens, we incorporate a time component in our interaction graph construction. For example, a CNOT within the earliest layer of the circuit that acts on the same qubit(s) as many later CNOTs (i.e. a CNOT that is ``blocking'' many other gates and acting as a bottleneck) should be given higher priority in qubit placement than a CNOT near the end of the circuit. In fact, CNOTs many layers back should not be considered at all, as they are ``shielded'' from affecting the current permutation we wish to generate by earlier CNOTs. Therefore, while we do look-ahead past the first layer, we limit how far we look and down-weight less important CNOTs. We mention at this point that limiting the look-ahead also has the added benefit of improving runtime; we further discuss this in Sec.~\ref{sec:runtime}. Finally, we must also consider the previous permutation we chose. Therefore, adjacency of the previously chosen permutation is incorporated into the interaction graph as well. 

After the creation of the interaction graph, we make use of spectral graph theory, as described in Sec.~\ref{sec:spectral}. Note that this gives a mapping of each logical qubit onto a coordinate location according to the graph Laplacian eigenvectors, which can then in turn be used to place the qubits at discrete integer locations. As we have already seen other algorithms make use of the MinLA problem~\cite{pedram2016layout,shafaei2013optimization} and graph partitioning~\cite{chakrabarti2011linear}, we believe the related spectral graph drawing method provides another good solving strategy to the qubit mapping problem. 

Like with the other algorithms that have used an interaction graph, our use of spectral graph theory leads us to focus on the linear nearest neighbor architecture. We do not see this as a prohibitive restriction on the problem; as described in Sec.~\ref{chap:one}, many experimental devices do in fact adhere to LNN connectivity. Furthermore, because LNN connectivity is among the most restrictive of architectures, finding an efficient translation to an LNN compliant circuit implies the original circuit can be run efficiently on many other practical, and often less restrictive, architectures~\cite{cheung2007translation}. Lastly, since many analyses have been done on LNN architectures and most qubit mapping algorithms can run with LNN connectivity constraints, comparison of our results to others can be used to specifically evaluate our solving strategy choices (i.e. using a weighted interaction graph and a spectral graph drawing mapper). These reasons all justify and motivate our focus on LNN architectures.

\subsection{Spectral Mapper}\label{sec:spec-mapper}
In this section, we formally detail our spectral mapper.

\subsubsection{CNOT Dependency and Layering}\label{sec:cnot-dep}
Let $C$ be the CNOTs of a logical circuit. The CNOTs of $C$ have a dependency in that, for $i < j$, CNOT $j$ must be applied strictly later than CNOT $i$ if there is a qubit both CNOTs act on. In other words, CNOT $j$ cannot be commutated ahead of CNOT $i$. We say that CNOT $i$ is a \textit{direct blocker} of CNOT $j$ if:
\begin{enumerate}
	\item $i < j$,
	\item they both act on qubit $q_k$, and
	\item no CNOT between $i$ and $j$ acts on $q_k$.
\end{enumerate}
Let $b\left(j\right)$ be the set of direct blockers of CNOT $j$. In general, $b\left(j\right)$ can have $0, 1$ or $2$ elements.

Next, we define $t_j^f$, the \textit{forward layer} of CNOT $j$. It is a quantity that represents the minimum number of CNOT layers that must be applied before CNOT $j$ is eligible to be applied. It acts as a proxy for how ``soon'' the CNOT can be applied. Formally, we have:
\begin{equation}
t_j^f = \begin{cases}
0, & \textrm{if }\lvert b\left(j\right)\rvert = 0\\
1 + \displaystyle\max_{i\in b\left(j\right)} t_i^f, & \text{otherwise}
\end{cases}.
\end{equation}
If we think about assigning each CNOT a layer, where each layer contains CNOTs that can be simultaneously applied, $t_j^f$ represents the layer of CNOT $j$ in a greedy layering that tries to place each CNOT as early as possible. If we consider all the CNOTs $j$ for which $t_j^f = 0$, one of them will be the next CNOT to be applied. We will call this set of CNOTs the \textit{front layer}.

Consider now the reverse list of $C$, given by $\left[\left(l_N^c, l_N^t\right), \ldots, \left(l_1^c, l_1^t\right)\right]$. We similarly define $t_j^r$, the \textit{reverse layer} of CNOT $j$, as the forward layer of CNOT $j$ in this reversed list. Let $T = \max_i t_i^f$ be the maximum forward layer. Then, the quantity $T - t_j^t$ represents the layer of CNOT $j$ in a lazy layering of $C$ that tries to place each CNOT as \textit{late} as possible. 

We interpret these results as follows. A small forward layer means the CNOT is soon to be eligible for application, and has the potential to be related to the next mapping of logical to physical qubits. A simultaneously small reverse layer, however, means that the CNOT could in fact be applied much later, and have little to do with the next mapping. These concepts are used when calculating our weighted interaction graph.

\subsubsection{Weighted Interaction Graph}
A weighted interaction graph is some weighted, undirected graph $\Tilde{G} = \left(Q, \Tilde{E}\right)$ whose vertices are logical qubits, and whose edge weights $w_{ij}$ represent the priority of placing those qubits close to one another. There are two categories of contributions to edge weights: the previous logical to physical mapping, and the CNOTs yet to be applied. We consider each one.

First, suppose the previous permutation is given by $\pi: Q\rightarrow V$. For simplicity, we will let $V = \left\{1, 2, \ldots, M\right\}$, and order the vertices so that the LNN connectivity constraint places consecutive numbers adjacent to each other. That is, for $i, j\in V$,
\begin{equation}
\left(i, j\right)\in E \text{ if and only if } \left\lvert i - j\right\rvert = 1.
\end{equation}
Then, for every $q_i, q_j\in Q$ for which $\left(\pi\left(q_i\right), \pi\left(q_j\right)\right)\in E$, i.e. for each pair of logical qubits that were previously adjacent, we increment $w_{ij}$ by the \textit{prior mapping weight} $\beta\in \left[0, 1\right]$. The motivation is that, to minimize the number of SWAPs, the next permutation we generate should be as close to the previous as possible. The parameter then acts as a sort of memory for the algorithm. It adds a $\beta$-weighted component of the previous logical adjacency to $\Tilde{G}$.

Next, we consider the unapplied CNOTs, which we will represent as $C$. Each CNOT can only be applied if the qubits they act on are adjacent. Furthermore, at least one CNOT should be applied in the next generated permutation, and at least one must come from the front layer. As reasoned in Secs.~\ref{sec:design} and~\ref{sec:cnot-dep}, we can consider CNOTs with a small forward layer to potentially be applied in the next mapping, while discounting those same CNOTs with large backward layers due to the flexibility in their layer choice. We define an integer $\tau \in \left[0, T\right]$ to be the \textit{cutoff layer} and real number $\alpha\in \left[0,1\right]$ to be the \textit{layering discount}. Then, if CNOT $i$ acts on qubits $q_j$ and $q_k$ and has forward layer $t_i^f\leq \tau$, we increment $w_{jk}$ by $\alpha^{T - t_i^r}$.

In closed form, we define the edge weight between $q_j$ and $q_k$ as:
\begin{equation}
w_{jk} = \sum_{\substack{i: \left\{l_i^c, l_i^t\right\} = \left\{q_j, q_k\right\}\\t_i^f\leq \tau }} \alpha^{T - t^r_i} + \begin{cases}
\beta & \text{ if } \left(\pi\left(q_j\right), \pi\left(q_k\right)\right)\in E\\
0 &\text{otherwise}
\end{cases}.
\end{equation}

We intend to use this interaction graph $\Tilde{G}$ as an input to the spectral drawing method outlined in Section~\ref{sec:spectral}. A mapping $\pi: Q\rightarrow V$ is generated, placing the logical qubits at physical locations. While we have chosen the edge weights of $\Tilde{G}$ to try to place the qubit pairs in front layer CNOTs close together, there is actually no guarantee that any front layer CNOTs are actually compliant on the generated mapping. This leads to a breakdown that needs to be resolved via some fallback. A naive fallback would be to arbitrarily pick a front layer CNOT and SWAP the two qubits together, or to apply another established algorithms. In the following section, we propose a third strategy, which we ultimately use.

\subsubsection{Forced Coupling}
For our fallback strategy, we would like to make use of our weighted interaction graph  and spectral drawing framework. However, we wish to guarantee that the generated permutation allows for at least one front layer CNOT to be compliant. To this end, we propose a \textit{forced coupling} step. First, we consider the front layer CNOTs with the largest reverse layers, signifying the highest priority of application within our previous description. We use this limited subset of the front layer, rather than the whole front layer, to minimize the impact of the fallback and only apply this forceful coupling to priority pairs of qubits. 

More specifically, we are seeking every CNOT $i$ for which $t_i^r = T$. Then, if that CNOT $i$ is applied on qubits $q_j$ and $q_k$, we combine the corresponding vertices in $\Tilde{G}$ together. This combination involves:
\begin{enumerate}
	\item replacing $q_j$ and $q_k$ with a single node fused node $f$ representing the two qubits, and
	\item for every third node $p$ (which may be a single qubit or a newly created fused node), drawing an edge between $f$ and $p$ with edge weight equal to the sum of the previous edge weights between $q_j$ and $p$, and between $q_k$ and $p$.
\end{enumerate}
After this process is performed for every CNOT $i$ with $t_i^r=T$, we are left with a new interaction graph $\Tilde{G}^{forced} = \left(Q^{forced}, \Tilde{E}^{forced}\right)$.

Note that because no front layer CNOTs act on the same qubit, each new vertex in $Q^{forced}$ represents either one or two of the original qubits from $Q$. We then again use the spectral drawing method, but with $\Tilde{G}^{forced}$ as the input. 

This method will first provide us with a real number coordinate position for each vertex in $Q^{forced}$. For each coordinate, we first add a small perturbation, randomly generated for each vertex, so as to break ties; we do this to guarantee the forced pairs will be adjacent to each other. Then, we assign each of the original qubits from $Q$ to the coordinate of their corresponding forced vertex in $Q^{forced}$; note that forced pairs are thus given the same location. Finally, if $y_i$ is the coordinate for $q_i$, we generate the permutation $\pi: Q \rightarrow V$ adhering to
\begin{align}\label{eq:forced-ordering}
\pi\left(q_j\right) > \pi\left(q_k\right)\implies y_j\geq y_k.
\end{align}
Note that the ties between the forced pairs is broken at random, as any such ordering of the pairs will satisfy Eq.~\ref{eq:forced-ordering}.

By virtue of the forced pairs selection, at least one of the front layer CNOTs is connectivity compliant on $\pi$, and therefore the overall algorithm can progress.

One interesting consideration is whether this forced pairing algorithm can be used independently, as a standalone mapper, rather than as a fallback. This is an option we consider and allow for our overall algorithm.

\subsubsection{Applying the Spectral Drawing Algorithm}
We conclude our description of our spectral mapper by discussing some details regarding the use of the spectral drawing method from Sec.~\ref{sec:spectral}. As mentioned, the coordinates generated may sometimes lead to vertices assigned to the same location. In that case, ties are typically broken at random, either implicitly when moving the vertices to discrete integer locations, or explicitly through the addition of small, random perturbations to each coordinate. 

Another consideration we make is in regards to the symmetries of our architecture. In particular, a given mapping to an LNN architecture affords the same connectivity constraints when rotated $180^{\circ}$ (or reflected about its center). As a result, we must consider both possibilities; this is equivalent to considering the mapping induced by both the Fiedler vector $y^{(2)}$ calculated by the eigensolver, and its negative $-y^{(2)}$. The mapping we ultimately choose is the one which requires the fewest SWAPs to get from the previous permutation; as a proxy, we use as a metric the sum of the distances each logical qubit must travel from previous to potential mapping, and pick the potential mapping with the smaller sum as the generated mapping.

\subsection{Bidirectionality}
Many of the algorithms we have considered start from the beginning of the circuit, as it is provided, and then sequentially work to the end in the forward direction. When considering the connectivity compliance problem on a circuit of CNOTs, however, the same problem can be solved on the reversed circuit to provide a valid transformation. Indeed, there is no inherent preference between the two directions. Thus, we explore the option of a bidirectional mapper. At each iteration, the mapping strategy is applied as described in Sec.~\ref{sec:spec-mapper} to generate a mapping $\pi$ that allows for some front layer CNOTs to be applied. Our bidirectional proposal adds for the same iteration the generation of a mapping on the reversed circuit as well, so that some front layer CNOTs on the reversed circuit can be applied. In terms of the notation for generating the weighted dependency graph, the reverse mapper simply interchanges $t_i^f$ and $t_i^r$. The mappings are then generated for each end of the circuit every iteration, working towards the middle of the circuit.

\subsection{Overall Connectivity-Compliance Transformation Algorithm}
In this section, we summarize the full algorithm for transforming a logical circuit into one that is LNN compliant. First, we list the configuration options we have available for our algorithm. These include whether to use the forced-coupling mapper as a standalone or fallback strategy; whether to include bidirectionality; and what choices of $\alpha, \beta$, and $\tau$ to choose. A summary of the options is given in Table~\ref{tab:configurations}.

\begin{table}[h!]
	\centering
	\begin{tabular}{||p{3cm} || p{7cm}|p{3cm}||} 
		\hline
		\textbf{Parameter} & \textbf{Description} & \textbf{Values}\\ [0.5ex] 
		\hline\hline
		Forced-coupling & Whether the mappings are exclusively based on the forced-coupling weighted interaction graph, or if the forced-coupling case is only used as a fallback for the regular weighted interaction graph. & \{Standalone, Fallback\} \\ 
		\hline
		Direction & Which direction(s) the iterations traverse through the circuit. & \{Forward, Bidirectional\} \\
		\hline
		$\alpha$ & The layering discount factor used in calculating the weighted interaction graph. It is meant to diminish the impact of CNOTs that can be pushed back to much later layers. & Any real in $\left[0,1\right]$ \\
		\hline
		$\beta$ & The prior mapping weight used in calculating the weighted interaction graph. It is meant to weight the relative importance of remaining close to the previous mapping. & Any real in $\left[0,1\right]$ \\
		\hline
		$\tau$ & The cutoff layer used in calculating the weighted interaction graph. It is meant to limit the layers of look-ahead. & Any positive integer\\
		\hline
	\end{tabular}
	\caption{A summary of all possible configuration options for our overall circuit transformation algorithm.}
	\label{tab:configurations}
\end{table}

We begin with a logical circuit $C$, consisting of $N$ CNOT gates and $M$ qubits. Until we run out of unapplied CNOTs, we perform the following iteration. First, we run our spectral mapper, as described in Sec.~\ref{sec:spec-mapper}, using our specified configurations. This involves the creation of a weighted interaction graph, the use of spectral graph theory to label each logical qubit to a coordinate, and the ordering of logical qubits into discrete locations on an LNN architecture. Note that for the first iteration, no previous mapping is used for the interaction graph. By design, some front layer CNOTs are connectivity compliant on the generated mapping. All CNOTs that we can apply to this mapping are applied and removed from the circuit. Note that as CNOTs are applied, we update the layerings and try to apply any CNOTs that newly enter the front layer. Once a front layer is reached with no connectivity compliant CNOTs, the current iteration is over. Note that for the bidirectional configuration, the same operations are also done but from the end of the circuit with the reversed circuit mapping generated.

After all CNOTs are applied, we have a sequence of permutations, with CNOTs applied during each permutation. The final step is to use the token swapping permuter from Sec.~\ref{sec:token} to generate a list of SWAPs needed to transition between successive permutations. The resulting circuit, which contains alternating layers of CNOTs compliant on the same permutation and SWAPs moving the logical qubits between permutations, is our final, equivalent LNN compliant circuit.

\subsubsection{Special Implementation Details}\label{sec:details}
To close out the description of our algorithm, there are some specific implementation details that we wish to describe. The first is related to organization of the unapplied CNOTs. Note that it is important to be able to determine the forward and backward layerings for each CNOT, to determine which CNOTs are part of the front layer (and are therefore eligible for application), and to update the layerings after each iteration of CNOT application. We do this by keeping a dependency graph, where the nodes are each CNOT and the edges connect each CNOT to its direct blocker. We augment this graph with layer information at each node, which can be done by traversing the CNOTs in order and using the blocking information to sequentially determine the layering. We also keep track of a mapper from each layer to the CNOTs contained, for quick reference. When a permutation is generated, we can then reference the front layer with ease and consider the compliancy of each. When a compliant front layer CNOT is found, it is deemed applied and removed from the set of unapplied CNOTs; the CNOTs which were directly blocked are then considered for front layer status. Once all front layer CNOTs are no longer compliant for the permutation, a single pass is made through the remaining CNOTs to reupdate dependencies and layering information.

The second detail is categorically different, and is more numerical in nature. When we seek the Fiedler vector of the Laplacian matrix $L$, there are some properties of our problem that allow for more efficient calculation. First, because $L$ is a real and symmetric matrix, special eigensolvers, like the Lanczos method, can take advantage of the structure. Furthermore, iterative methods, like the Lanczos method, allow for the calculation of just a few eigenvectors in many fewer operations than full eigensolvers that determine the entire spectrum. Finally, because we ultimately only care about the order of the components of the eigenvector, the precision is not ultimately that important; consequently, the number of eigensolver iterations can also be minimized.

\subsection{Runtime Analysis}\label{sec:runtime}
In this section, we analyze the runtime of our algorithm. The number of mapping iterations is $O\left(N\right)$ where $N$ is the number of CNOTs in the circuit, as every iteration is guaranteed to apply at least one CNOT\@. Each time the mapper is used, we need to construct $\Tilde{G}$, find an eigenvector of the corresponding $M\times M$ Laplacian (where $M$ is the number of qubits in the circuit), apply front layer CNOTs, and update the dependency graphs. The calculation of $\Tilde{G}$ requires $O\left(N\right)$ weight calculations, dependent on how many CNOTs fall within the cutoff layer $\tau$. Next, we bound the calculation time of the Fiedler vector to be $O\left(M^3\right)$. It is shown by~\cite{pan1999complexity} 
that the runtime of calculating the eigenvalues and eigenvectors of an $M\times M$ matrix, to a relative error of $O\left(2^{-b}\right)$, is bounded by $O\left(M^3 + M\log^2\left(M\right)\log\left(b\right)\right)$. As mentioned in Sec.~\ref{sec:details}, because we only need a rough estimate of the Fiedler vector, $b$ need not be large. In practice, the runtime is typically bounded by the $O\left(M^3\right)$ term. Second, we note that in practice we only require a few iterations of an iterative eigensolvers, like the Lanczos method, which can often compute the desired eigenvector in many fewer operations than calculating the full spectrum for a general matrix.

After the mapping is generated and CNOTs are applied, a single traversal of the unapplied CNOTs is needed to update the layerings; this takes another $O\left(N\right)$ operations. Each mapping generated will need to be permuted to the next via the approximate token swapping algorithm; from~\cite{childs2019circuit,miltzow2016approximation}, we can determine that at most $O\left(M^2\right)$ steps of the algorithm is needed, with each traversal of the companion graph $F$ requiring $O\left(M\right)$ time. Overall, one call to the token swapping algorithm therefore takes $O\left(M^3\right)$ time.

Therefore, for each of the $O\left(N\right)$ permutations we generate, we require $O\left(N+M^3\right)$ time. The overall runtime for transforming a single circuit is therefore $O\left(N^2+NM^3\right)$. In practice, $M \ll N$ (usually circuits are large in gate count and while the number of qubits is limited). Consequently, the scaling is dominated by $O\left(N^2\right)$.

\subsection{Choosing Optimal Configurations}

Because our algorithm has many possible configuration options, we started by evaluating our algorithm on a wide set of options. We considered whether or not to allow the mapping to occur bidirectionally; whether the forced pairing of first layer qubits was used standalone or only as a fallback; and allowed each of the weighting (both for future CNOTs discount $\alpha$ and for the previous permutation weight $\beta$) to take on values from 0.1 to 0.9 (inclusive), in intervals of 0.1. This totaled 384 configurations. Note that we fixed the cutoff layer to $\tau = M$ for the regular weighted interaction graph calculations, and $\tau = 4M$ for the forced pairs calculations. The purpose is to provide enough layers so that the resulting interaction graph contained all qubits.

We tested these configurations over our test set (described below in \Cref{chap:four}) and found that for the vast majority of the benchmarks, the best performing configuration used the forward direction only and used forced pairing as a fallback. This allows us to narrow down our choices for those two parameters.

Next, for just the forward-direction, forced pair fallback results, we considered each pair of weights $\left(\alpha, \beta\right)$. For each pair $\left(\alpha, \beta\right)$, we counted the number of benchmarks for which the resulting number of added SWAPs was within 5\% of the minimum number of added SWAPs across all pairs of $\left(\alpha, \beta\right)$. The $\left(\alpha, \beta\right)$ that achieved the highest count was chosen, the benchmarks which contributed to that pair's count was discarded, and the process was repeated until a total of ten $\left(\alpha, \beta\right)$ pairs were generated. These ten pairs were found to be:
\begin{align*}
\{&\left(0.2, 0.3\right), \left(0.3, 0.4\right), \left(0.4, 0.1\right), \left(0.5, 0.1\right), \left(0.5, 0.6\right), \\
&\left(0.7, 0.1\right), \left(0.8, 0.1\right), \left(0.8, 0.2\right), \left(0.8, 0.6\right), \left(0.9, 0.9\right)\}.
\end{align*}

The method by which these resulting ten pairs are chosen is motivated by our desire to have a relatively good coverage by producing near best results for all of the benchmarks. 

Therefore, rather than just running our circuit transformation algorithm with just one pair of $\left(\alpha, \beta\right)$ weights, our overall ``meta-algorithm'' runs the algorithm for each of these ten configurations on a given circuit. The result with the smallest number of added SWAP gates is then returned as the transformed circuit.

\subsection{Software Implementation}
We implemented our algorithm in Python 3. The source code can be found at \url{https://github.com/joelin0/spectral-mapping}. Note that the code is subject to change, so the repository README and source code itself provide the most up-to-date information regarding the algorithm implementation. 

\section{Performance Testing Methodology}\label{chap:four}
\subsection{Benchmarks}
To test our algorithm's performance, we use benchmarks that resemble realistic circuits that may be seen in practice. The set of benchmarks we primarily focused on are publicly available OpenQASM benchmarks used by Zulehner et al.~\cite{zulehner2018efficient} to evaluate their own algorithm; these files can be found at~\url{https://github.com/iic-jku/ibm_qx_mapping/tree/master/examples}. Most of these circuits come from RevLib~\cite{WGT+:2008}, a database of reversible and quantum circuit benchmarks. Note that others who have also approached the problem of circuit connectivity-compliance, like Cowtan et al.~\cite{cowtan2019qubit}, have used the same exact benchmark files to evaluate their algorithm. Other still, like~\cite{wille2014optimal} and~\cite{shafaei2013optimization}, have used RevLib circuits (although one should be cautious in comparing their results with those using these OpenQASM files, as the circuit decomposition used from RevLib circuit to quantum gates is not specified and may differ). 

The benchmarks we used have between 3 and 16 qubits and up to 10,000 CNOTs in their original circuit. Note that due to computational resource and time constraints, we were unable to evaluate larger benchmarks across all of the algorithms.

We also supplemented our testing with benchmarks used by~\cite{NRSCM}. In particular, we use some of the smaller, post-optimization Arithmetic and Toffoli benchmarks, which can be found at~\url{https://github.com/njross/optimizer/tree/master/Arithmetic_and_Toffoli}. These benchmarks have between 5 to 19 qubits and up to 130 CNOTs.

\subsection{Comparisons with Alternate Algorithms}
After running our algorithm, we compare its output with the algorithms of Childs, Schoute, and Unsal~\cite{childs2019circuit}, and Zulehner, Paler, and Wille~\cite{zulehner2018efficient}. For conciseness, we shall refer to these algorithms as CSU19 and ZPW18, respectively. We choose these two implementations, as they are among the most recently developed methods, were submissions for an IBM competition on this specific problem (albeit on IBM's architecture), can all operate on LNN architectures, and, most importantly, have open source code available. As of the writing of this thesis, the CSU19 source code can be found at~\url{https://gitlab.umiacs.umd.edu/amchilds/arct/tree/master} and the ZPW18 source code can be found at~\url{https://github.com/iic-jku/ibm_qx_mapping}.

Each tested code has a command line interface that allows for OpenQASM files to be passed in and for an equivalent LNN compliant circuit to be written out. Each code is modified to measure the total amount of CPU time needed to run the entire process. Note that the code from CSU19 was modified to accept arbitrary QASM files, and also to remove writes of files beside the final output QASM\@. Note further that the CSU19 algorithm also has four different choices of mappers to minimize added SWAP count (called greedy size, simple size, extension size, and qiskit size); we ran each one separately on the suite of benchmarks.

Each benchmark test was run on a device with 2 vCPUs and 4 GB of RAM\@; with a few exceptions, these constrained computational resources did not affect the benchmark testing.

\section{Benchmarking Results and Discussion}\label{chap:five}
Now, we present the results of the benchmark testing described in Sec.~\ref{chap:four}, and compare the performance of each algorithm. We refer to our algorithm as the Spectral algorithm; the four size-optimizing configurations of~\cite{childs2019circuit} as CSU19 greedy size, simple size, extend (extension) size, and qiskit-based; and the algorithm from~\cite{zulehner2018efficient} as ZPW18. 

The raw data for the experiments is given in Table~\ref{tab:data} of Appendix~\ref{app:raw-data}. There, we present each benchmark circuit's name, number of qubits ($M$), and number of CNOTs ($N$). For each algorithm, we provide the number of SWAP gates in the generated connectivity compliant circuit of that benchmark, as well as the CPU time taken. Note that we first list the selected benchmarks used by~\cite{zulehner2018efficient}, and below them the selected benchmarks from~\cite{NRSCM}.

\begin{figure}[!h]
	\includegraphics[width=\textwidth]{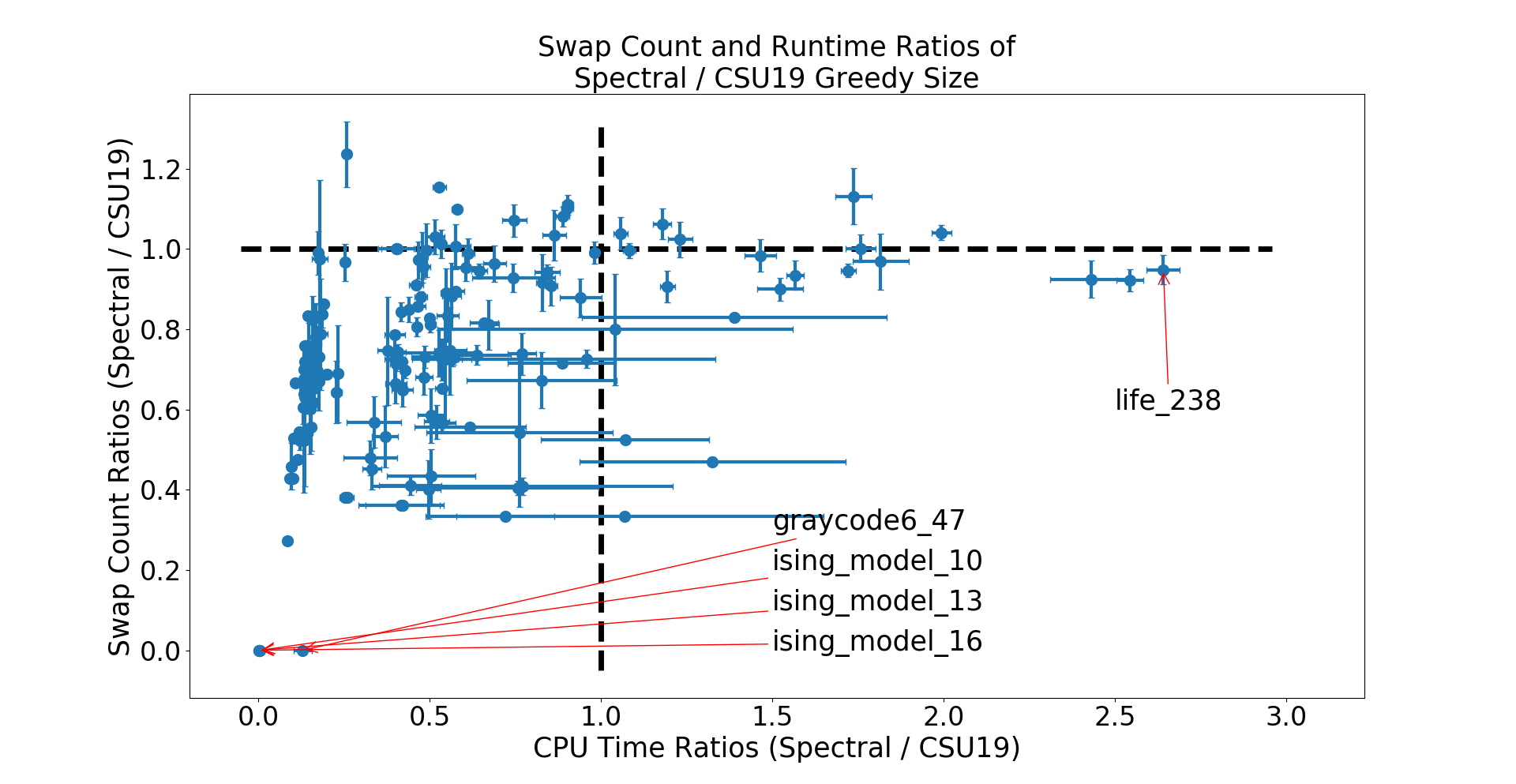}
	\caption[Performance Ratios for Spectral Algorithm vs. CSU19 Greedy Size Algorithm]{Plot of swap count ratios vs. CPU runtime ratios for our spectral algorithm over the CSU19 greedy size algorithm. Each point represents a single benchmark test.}
	\label{fig:spect-vs-arct-greedy}
\end{figure}

\begin{figure}[!h]
	\includegraphics[width=\textwidth]{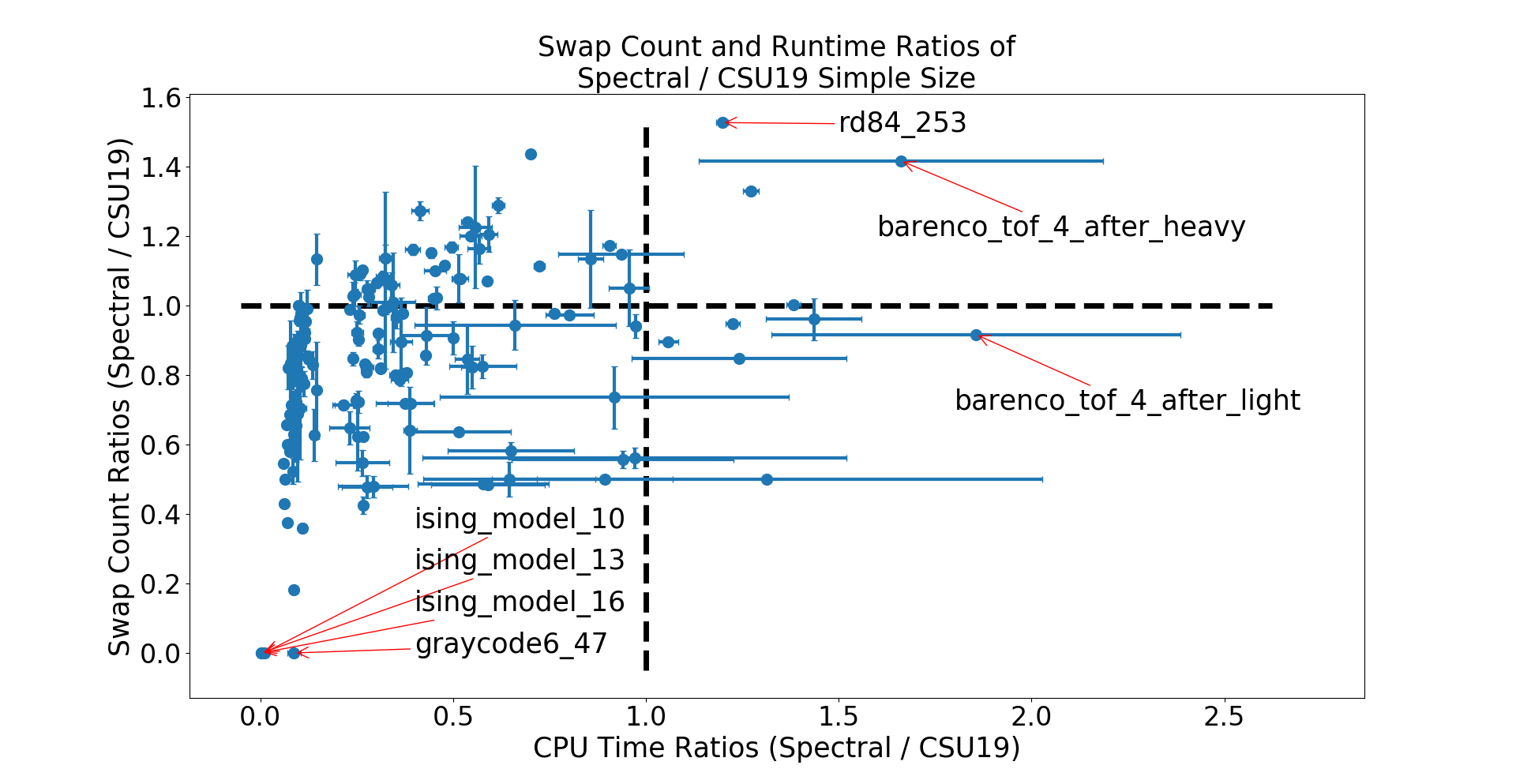}
	\caption[Performance Ratios for Spectral Algorithm vs. CSU19 Simple Size Algorithm]{Plot of swap count ratios vs. CPU runtime ratios for our spectral algorithm over the CSU19 simple size algorithm. Each point represents a single benchmark test.}
	\label{fig:spect-vs-arct-simple}
\end{figure}

\begin{figure}[!h]
	\includegraphics[width=\textwidth]{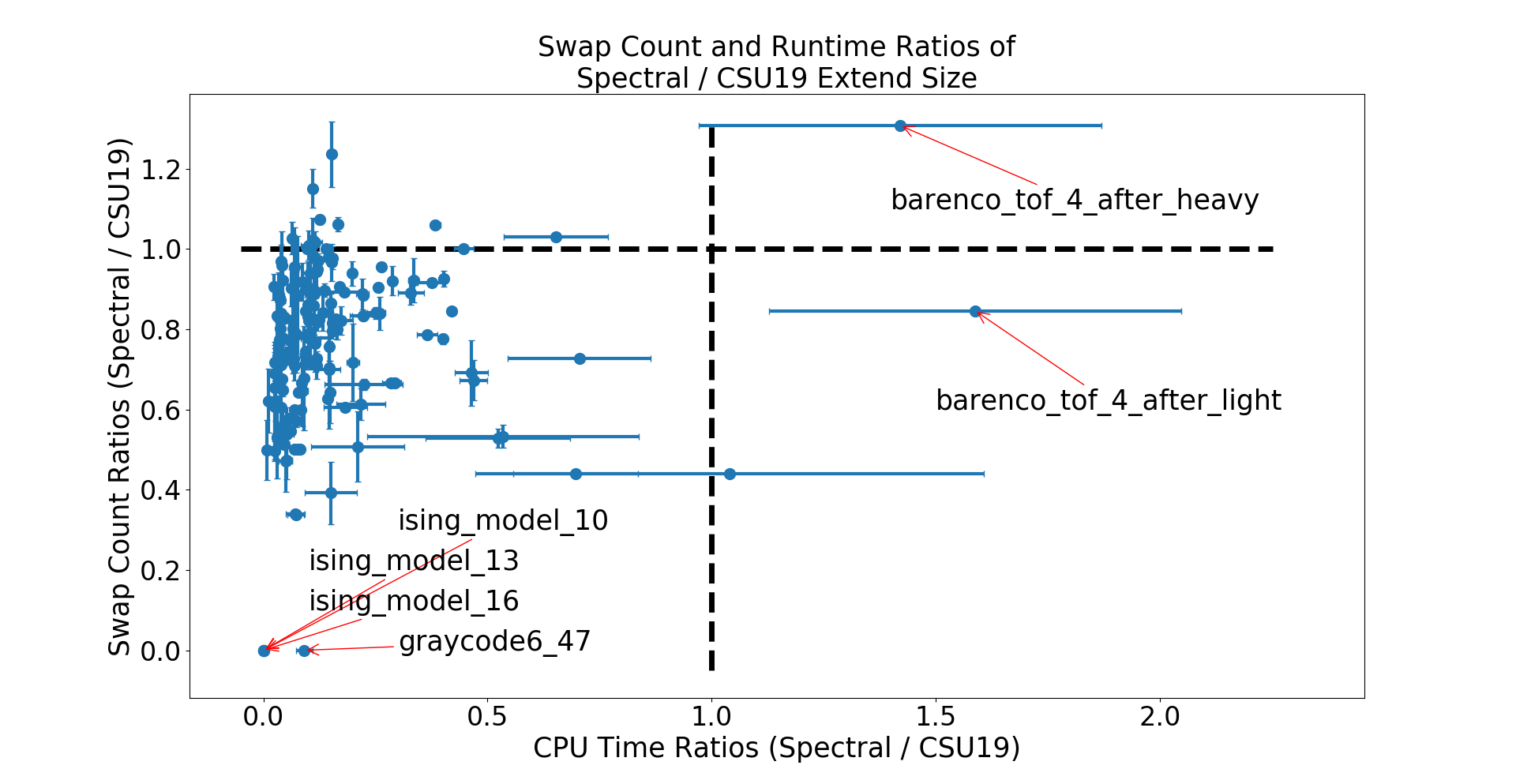}
	\caption[Performance Ratios for Spectral Algorithm vs. CSU19 Extension Size Algorithm]{Plot of swap count ratios vs. CPU runtime ratios for our spectral algorithm over the CSU19 extension size algorithm. Each point represents a single benchmark test.}
	\label{fig:spect-vs-arct-extend}
\end{figure}

\begin{figure}[!h]
	\includegraphics[width=\textwidth]{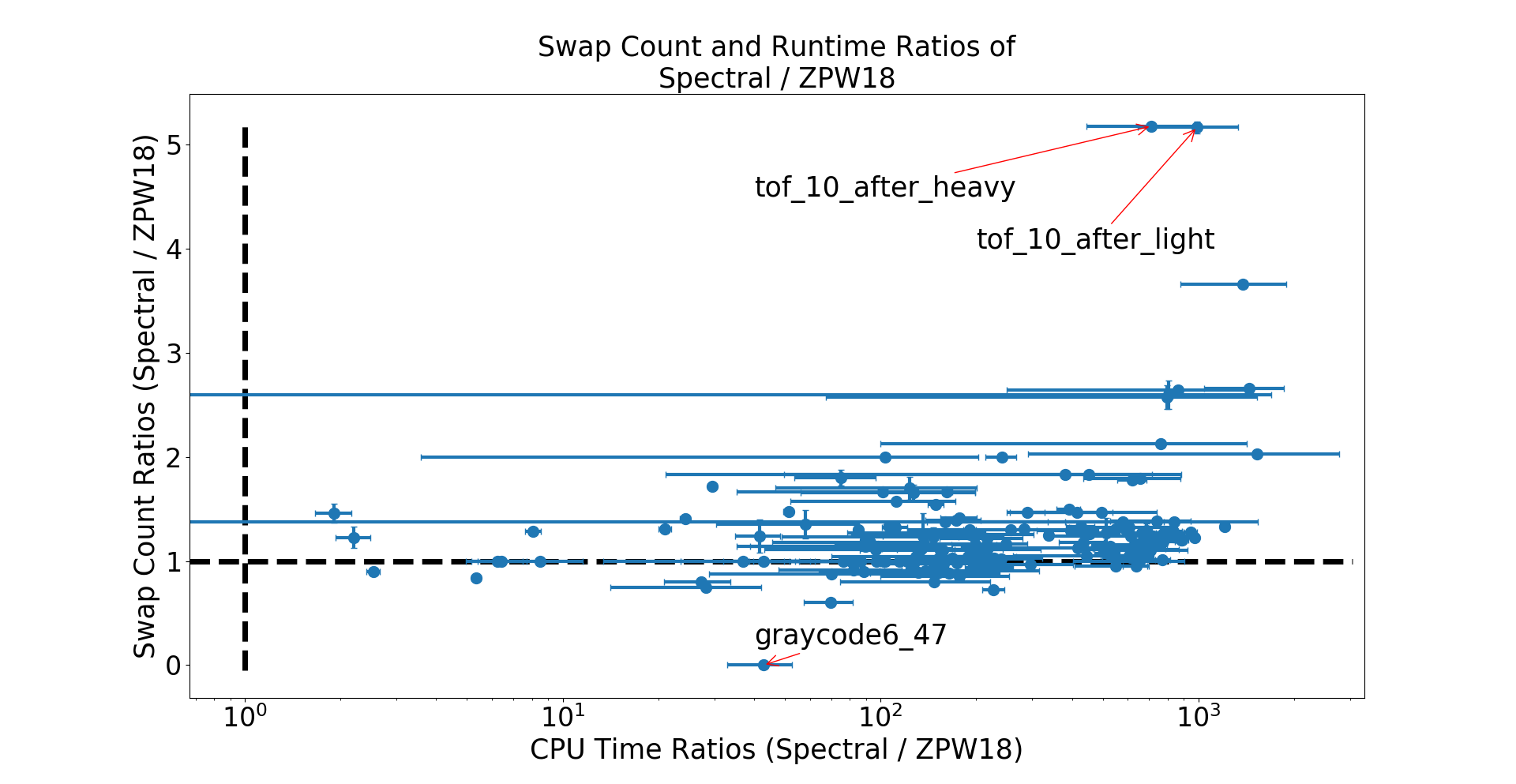}
	\caption[Performance Ratios for Spectral Algorithm vs. ZPW18 Algorithm]{Plot of swap count ratios vs. CPU runtime ratios for our spectral algorithm over the ZPW18 algorithm. Each point represents a single benchmark test.}
	\label{fig:spect-vs-ibm-qx}
\end{figure}

For the purpose of visual comparison, we provide plots in Figs.~\ref{fig:spect-vs-arct-greedy},~\ref{fig:spect-vs-arct-simple},~\ref{fig:spect-vs-arct-extend},~\ref{fig:spect-vs-arct-simple}, and~\ref{fig:spect-vs-ibm-qx}. These plots compare the results of our algorithm to each alternate algorithm: on the horizontal axis is the ratio of our CPU time taken to the alternate algorithm, while the vertical axis is the ratio of our added SWAP count to the alternate algorithm. A point is added for each benchmark result. If we draw lines at the ratio $1$ for both axis, four quadrants are created. The bottom left quadrant signifies that our algorithm produces a better circuit in faster time, and is the location we wish for most points to lay. Conversely those in the upper right quadrant represent benchmarks for which our algorithm does poorer in both performance metrics. For each plot, we also label some of the extreme points, some of which we analyze in further detail later in this Section.

\begin{figure}[!h]
	\includegraphics[width=\textwidth]{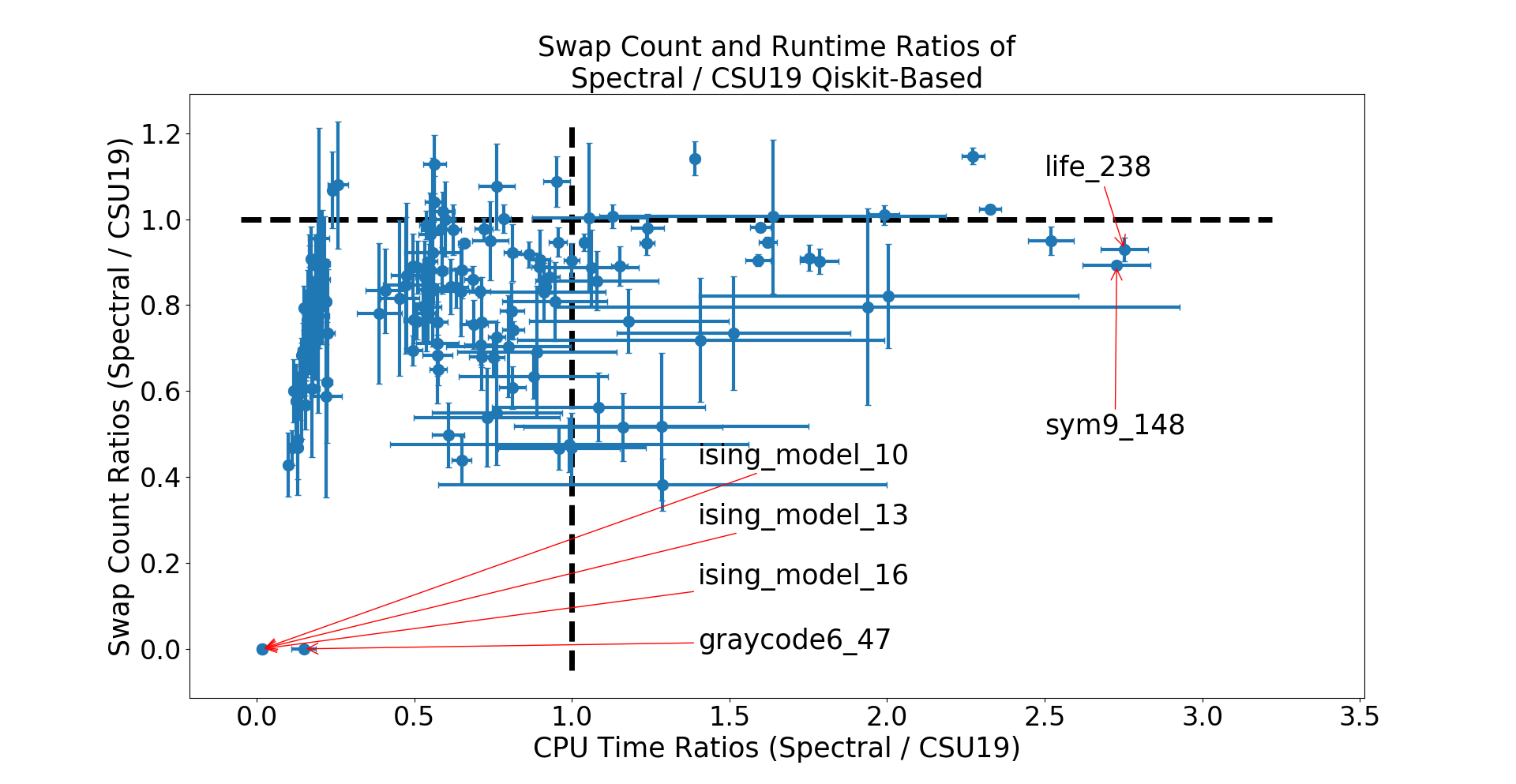}
	\caption[Performance Ratios for Spectral Algorithm vs. CSU19 Qiskit Size Algorithm]{Plot of swap count ratios vs. CPU runtime ratios for our spectral algorithm over the CSU19 qiskit-based algorithm. Each point represents a single benchmark test.}
	\label{fig:spect-vs-arct-qiskit}
\end{figure}

\subsection{Comparisons with CSU19 Algorithms}
First, we compare our algorithm to the CSU19 algorithms. From a design and implementation standpoint, the algorithms share much in common. Our algorithm is modeled in part after the mapper-permuter model of the CSU19 algorithms. Both algorithms use the approximate token swapping algorithm as the permuter. And, importantly from an experimental standpoint, both are implemented in Python. The major difference is in the mapper: we use our spectral drawing method on a weighted interaction graph, while they solve an optimization problem over the selection of a first layer CNOT and of a permutation on which that CNOT is compliant. Recall that our algorithm has time complexity $O\left(N^2 + NM^3\right)$. On LNN architectures, the CSU19 greedy size, simple size, extension size, and qiskit-based algorithms have time complexities $O\left(NM^5\right)$, $O\left(NM^4\right)$, $O\left(N^2M^4\right)$, and $O\left(NM^3\right)$, respectively. Assuming $M \ll N$, which is the case for most of the benchmarks, we see that the extension algorithm run in $O\left(N^2\right)$ while the greedy, simple, and qiskit-based algorithms all run in $O\left(N\right)$ time. Therefore, we would theoretically expect, in the limit of large circuit gate counts $N$, to have comparable runtimes with the extension algorithm while having slower runtimes compared to the other three.

Looking at the results, we see that our algorithm performs better on both metrics for the vast majority of benchmarks. Against the extension algorithm, our algorithm ran faster on all benchmarks. Furthermore, our algorithm generated a smaller SWAP count graph in almost all benchmarks, with the few exceptions only being slightly higher in comparison. Against the greedy and qiskit-based algorithms, our SWAP counts were again better for almost all benchmarks and only slightly more for the ones where our algorithm performed worse. As expected from the runtime analyses, the qiskit-based algorithm is asymptotically the fastest, even when $M$ is factored in; indeed, we start seeing some larger benchmarks for which our algorithm performs up to three times as slowly. The greedy algorithm also has a better runtime $N$ dependence compared to ours, thus explaining the right trail of higher time ratios. Finally, our algorithm had the most mixed performance against the simple size algorithm. While the majority of points still lay in the lower left quadrant, our algorithm provided a circuit with more SWAPs in many more benchmarks. Furthermore, our worst benchmarks resulted in SWAP counts over 1.6 times that of the simple size algorithm.

Our overall conclusion is that our algorithm is quite competitive with the CSU19 algorithms. In most cases, our algorithm finds circuits with significantly fewer SWAPs. This may be expected at least in part due to our algorithm being more tailored towards LNN architectures through the use of spectral drawing, while their algorithms target arbitrary connectivities. In terms of runtime, our algorithm can be quite a bit faster for some smaller circuits, but in general is on the same order of magnitude in terms of runtime. Theoretically, we do expect our algorithm to get slower for much larger circuits; however, that limit is clearly not yet reached for circuits with up to 10,000 CNOTs\@. Even above those counts, our algorithm still has polynomial asymptotic runtime.

\subsection{Comparisons with ZPW18 Algorithm}
Next, we compare our algorithm to the ZPW18 algorithm. Recall that the ZPW18 algorithm involves an A* search across a state space involving all possible mappings. As there are $M!$ possible mappings, both worst-case runtime and storage complexities are $O\left(M!\right)$. However, with A* search, the actual runtime may be a lot less depending on the choice of heuristic; as a result, it is more difficult to translate these theoretical runtime characteristics into experiments. Another difficulty in the runtime comparisons has to do with implementation: the implementation of the ZPW18 algorithm we compare against was written in C++. That means their code was compiled first before run on the benchmarks, an advantage not afforded to either the CSU19 algorithm or ours.

When considering the results presented in Fig.~\ref{fig:spect-vs-ibm-qx}, we immediately see their runtimes are many orders of magnitude faster than ours. We believe that much of this is due to the difference in implementation language (C++ vs. Python); however, without testing, this is merely speculation. Our algorithm also only performs better on the SWAP count metric on a minority of the benchmarks; additionally, with only one exception, the SWAP count decrease is only modest. In most cases, our algorithm performs worse, hovering up to (and a few times beyond) 50\% more SWAPs.

Because this algorithm is relatively exhaustive in nature and is also regarded as one of the best performing algorithms for this problem, we find it encouraging that our algorithm does still find smaller SWAP count circuits in a nontrivial number of benchmarks. Additionally, we believe that our algorithm will scale better to larger circuits. In the benchmarks we tested, the qubit count $M$ only went as high as $16$; therefore, the runtime limitations of the algorithm did not yet show. However, as $M$ increases, we do expect the $O\left(M!\right)$ runtime to lead to the ZPW18 algorithm being infeasible. One scalability issue that already did arise was with memory. In four of the benchmarks, their algorithm ran out of memory. Again, with an $O\left(M!\right)$ worst case scaling for the space complexity, we would expect the memory needed to quickly rise to exorbitant levels. Indeed, this concern for scalability was also raised by the authors of the SABRE algorithm~\cite{li2019tackling}.

\subsection{Commentary on Specific Benchmarks}\label{sec:commentary}
We close this Section with commentary on some of the outlier benchmarks marked in the figures. We start with the \texttt{ising\_model\_10}, \texttt{ising\_model\_13}, \texttt{ising\_model\_16}, and \texttt{graycode6\_47} benchmarks. Each benchmark has the property that the circuit can be made LNN compliant without any additional SWAPs. In fact, the circuits are already written in an LNN fashion. Therefore, only a single mapping needs to be generated with no additional SWAPs. It can be seen that our algorithm indeed detects the structure in all of these circuits and reports an optimal, zero SWAP compliant circuit. This is not necessarily the case for any of the alternate algorithms, especially for \texttt{graycode6\_47}.

Many benchmarks on the far right of the figures, like \texttt{life\_238}, \texttt{sym9\_148}, and \texttt{rd84\_253} are among the largest circuits we consider in terms of gate count. As expected, the larger the $N$, the slower our algorithm becomes relative to those alternate algorithms.

One of our worst performing benchmark is \texttt{4mod5-bdd\_287}. The circuit itself is not that large (only 7 qubits and 31 CNOTs). However, the circuit properties do raise a weakness of our algorithm in its current form. After the first two gates are applied, two of the seven qubits are never operated on again. Rather than limiting ourselves to a five qubit subset of the LNN architecture, however, our mapper continues trying to map all seven qubits. This likely leads to the two unused qubits being disruptively placed in the middle of the circuit and moved around between mappings, rather than staying fixed on the end. Additionally, the CNOTs are all such that no CNOTs can be commutated past each other. This means the front layer is only a single CNOT. The fact that only one CNOT is eligible at a time could mean the number of iterations is quite high and could also mean increased reliance on the forced pairs fallback. All of these factors could explain why our algorithm takes an unusually long time on this benchmark yet still produces a relatively high SWAP count circuit.

Finally, \texttt{tof\_10\_after\_heavy} and \texttt{tof\_10\_after\_light} are two benchmarks with which the ZPW18 algorithm seems to vastly outperform the other algorithms. While the specific cause still requires more investigation, it can likely be attributed to the ZPW18 algorithm's ability to detect the ``staircase'' pattern of CNOTs in the two benchmarks.

\section{Conclusions }\label{chap:six}
\subsection{Summary and Significance}
We explored the use of spectral graph theory and drawing to map logical qubits to physical qubits in connectivity-constrained devices. Concerned with the problem of transforming logical circuits into connectivity-compliant ones, we first characterized the properties and strengths of many other modern algorithms and decided to follow a mapper-permuter model. We focused our exploration on linear nearest neighbor connectivities, and in particular explored how spectral drawing could be used as part of the mapper. We contributed a novel way of generating a more sophisticated weighted interaction graph based on the previously generated mapping and on layering properties of the unapplied CNOTs. Having presented a class of algorithms with several configuration options, we selected a set of ten options found to provide the best coverage across realistic benchmarks. The overall meta-algorithm then generated ten circuits using these different options and returning the best one. Finally, we compared our implementation to two recent, high-performing algorithms. We find our implementation to be quite promising, constructing circuits with smaller SWAP counts on many benchmarks while still having scalable space and runtime characteristics.

\subsection{Future Directions}
There are many future directions for this work. Some categories include:
\begin{enumerate}
	\item optimizing the current LNN algorithm,
	\item adapting the algorithm for other architectures and objective functions,
	\item making use of the spectral mapper in conjunction with other algorithms, and
	\item making use of other approximation or relaxation methods.
\end{enumerate}

There are many minor optimizations that can be made on the current algorithm to improve performance. As mentioned in Sec.~\ref{sec:commentary}, our algorithm does not properly detect when qubits go unused part way through iterations. One solution is to detect when this occurs and fix the unused qubits at the ends; then, continue the algorithm on just a subset of the physical qubits. More generally, further exploration of some common circuit patterns can help illuminate issues in our solver and allow us to optimize for them. Another improvement is to further optimize the interaction graph generation. To start off, a more exhaustive and systematic search over the parameters, (including some that were fixed, like $\tau$), may find more optimal parameters and also might elucidate why some parameters perform better than others.

We can also consider exploring variations of the connectivity-compliance problem. One obvious modification is to consider more general connectivities. Because our algorithm relies on spectral drawing, which involves mapping to coordinate locations and then to discrete locations, adapting to general connectivities would be quite challenging. However, a two-dimensional, nearest neighbors grid connectivity could be within the realm of possibility. Then, our algorithm would use not only the Fiedler vector, but also the eigenvector with the next smallest eigenvalue. These vectors then provide a coordinate location in 2D for each qubit. The question of mapping the coordinates to discrete grid locations is not as straightforward, as there is not a strict ordering of 2D coordinates. Even supposing a similar assignment strategy as in the 1D case were use, issues still arise related to a fallback strategy. It is nontrivial to decide how to place forced pairs: which orientation they are in, or which qubits to ``shift aside'' to make space. Additionally, when the number of physical qubits exceeds that of the logical qubits, it is nontrivial to decide which physical qubits go unused. All-in-all, the adaptation from LNN to a 2D grid still has many open questions. A more feasible modification to consider is a change in the objective. So far, we have focused on circuit size, via minimizing added SWAPs. However, another metric we can explore is minimizing circuit depth. This would require no change to the current algorithm initially and would involve measuring the circuit depth for the generated connectivity-complaint benchmark circuits. 

Third, we propose that ideas from our spectral mapper may be used in conjunction with other developed algorithms. Most straightforward is the inclusion of our spectral mapper as a mapper option in the CSU19 mapper-permuter algorithm framework, albeit only for LNN connectivities. There is also an opportunity to use our spectral mapper to provide an initial mapping. Many search-style algorithms, like the ZPW18 A* algorithm~\cite{zulehner2018efficient} and SABRE~\cite{li2019tackling}, need to be seeded with a good initial permutation; we suggest that our spectral mapper could be used for that purpose.

Finally, another direction we wish to explore is the search for different solving strategies. In this work, we proposed the use of spectral graph theory, a framework that was not previously explored for this problem. The motivation was that the spectral optimization problem provides a relaxation of the original problem, which is to determine a list of the best mappings. While the original problem cannot be exactly solved efficiently, the relaxed problem can. Right now, however, the relaxation is done mapping-by-mapping: though it has lookahead properties and considers the previous mapping, the optimization is still very much a local one. We leave as an open problem the search for another optimization problem or framework that is efficient to exactly solve and provides a \textit{global} relaxation. The desire would be to frame the problem in such a way that a \textit{single} problem can be solved, for all layers of CNOTs, and then the resulting relaxed solution can be translated (e.g. rounded, assigned to discrete locations) to provide a strong approximate solution to the original problem.
\begin{acks}
JXL was funded in part by NSF grant CCF-1729369. ERA was partially supported by a Lester Wolfe Fellowship and the Henry W.\ Kendall Fellowship Fund from M.I.T. AWH was funded by NSF grants CCF-1452616, CCF-1729369, PHY-1818914, ARO contract
W911NF-17-1-0433 and the MIT-IBM Watson AI Lab under the project {\it Machine Learning in
Hilbert space}.
\end{acks}


\clearpage
\newpage

\newcommand{\noopsort}[1]{} \newcommand{\printfirst}[2]{#1}
\newcommand{\singleletter}[1]{#1} \newcommand{\switchargs}[2]{#2#1}

\clearpage
\newpage
\appendix
\section{Spectral graph theory proofs}\label{app:spectral-proofs}
In this section we include proofs of the claims made in \Cref{sec:spectral}.  They are not original to our paper (see e.g.~\cite{koren2003spectral}) but are included here for convenience.

\begin{proof}[Proof of \Cref{lemma:lapl-quad}]
	With use of the fact that $w_{ij} = w_{ji}$ and $L_{ij} = L_{ji}$, we see that:
	\begin{align}
		x^T L x &= \sum_{i = 1}^n \sum_{j = 1}^n x_i x_j L_{ij}\\
		&= \sum_{i = 1}^n x_i^2 L_{ii} + 2\sum_{i < j} x_i x_j L_{ij}\\
		&= \sum_{i = 1}^n \sum_{k \neq i} x_i^2 w_{ik} - 2\sum_{i < j} x_{i} x_{j}w_{ij}\\
		&= \sum_{i < k} x_i^2 w_{ik} + \sum_{k < i} x_i^2 w_{ik} - 2\sum_{i < j} x_{i} x_{j} w_{ij}\\
		&= \sum_{i < j} x_i^2 w_{ij} + \sum_{i < j} x_j^2 w_{ji} - \sum_{i<j} 2x_ix_jw_{ij}\\
                &= \sum_{i<j} w_{ij}\left(x_i^2 - 2x_ix_j+x_j^2\right)\\
                &= \sum_{i < j} w_{ij} \left(x_i - x_j\right)^2
	\end{align}
	as desired.
\end{proof}

\begin{proof}[Proof of \cref{lemma:zero-eig}]
	\begin{align}
		(L1_n)_i &= \sum_{j = 1}^n L_{ij}\\
		&= \sum_{i \neq j} L_{ij} + L_{ii} \\
		&= \sum_{i \neq j} -w_{ij} + \sum_{k\neq i} w_{ik}\\
		&= 0.
	\end{align}
\end{proof}

\section{Approximate Token Swapping Algorithm}\label{app:token-swapping}
In this section, we outline an approximate token swapping algorithm as proposed in~\cite{miltzow2016approximation}. First, we define two operations. An \emph{unhappy swap} occurs when we perform a swap for which one token was already on its desired vertex and the other token is swapped closer to its desired vertex. A \emph{happy swap chain} occurs when, given a path of $l+1$ distinct vertices $v_{i_1}, \ldots, v_{i_{l+1}}$, we perform in order the $l$ swaps $\left(v_{i_1}, v_{i_2}\right), \left(v_{i_2}, v_{i_3}\right), \ldots, \left(v_{i_{l}}, v_{i_{l+1}}\right)$ and \emph{every} swapped token along the path is moved strictly closer to its target vertex.

It turns out that as long as some token is not at its desired vertex, one of the above operations exists. The approximate algorithm, then, involves finding either an \textit{unhappy swap} or a \textit{happy swap chain} and performing that operation. Furthermore, it was shown by~\cite{miltzow2016approximation} that this algorithm is guaranteed to converge.

To efficiently detect one of the two valid operations, a companion graph $F$ is created at each step. First, $F$ is given the same set of vertices $V$. For an edge between vertices $v$ and $w$ of the original graph $G$, consider a swap along that edge. If this causes the token currently on vertex $v$ to move closer to its desired vertex, then a \emph{directed} edge is added from $v$ to $w$ in $F$. A vertex with out-degree $0$ represents a token at its desired location, and therefore any edge going into that vertex represents an unhappy swap. Additionally, any directed cycle represents a happy swap chain. Therefore, if we start at any vertex whose token is misplaced and travel along a directed path, we will eventually end at a vertex with no outward vertices (thus detecting an unhappy swap) or return to a vertex already in the path (thus detecting an unhappy swap chain). This algorithm ultimately runs in time polynomial in the size of $G$, i.e. polynomial in $\left\lvert V\right\rvert$ and $\left\lvert E\right\rvert$. 

\section{Raw Data}\label{app:raw-data}
\pagestyle{empty}
In this section, we present the full results of our benchmarking experiments, described in Sec.~\ref{chap:five}. They appear in Table~\ref{tab:data}, across the next several pages.

\begin{landscape}
	
	\tiny
	\let\center\empty
	\let\endcenter\relax
	\centering
	\LTcapwidth=\textwidth

\begin{longtable}{|p{2.5cm}||p{0.7cm}|p{0.7cm}||p{1cm}|p{1cm}||p{1cm}|p{1cm}||p{1cm}|p{1cm}||p{1cm}|p{1cm}||p{1cm}|p{1cm}||p{1cm}|p{1cm}||p{1cm}|p{1cm}||p{1cm}|p{1cm}|}
	\caption[Results of benchmarking experiments on LNN architectures.]{Results of benchmarking experiments on LNN architectures. Each row is a single benchmark, where its name is the filename of the test circuit. Each benchmark is characterized in part by the number of qubits and number of CNOTs in the original circuit. The algorithms we compare are our spectral mapper, the four size mappers from CSU19~\cite{childs2019circuit}, and ZPW18 algorithm~\cite{zulehner2018efficient}. We run each algorithm, and characterize the performance both by the number of added SWAPs in the connectivity-compliant circuit, and the amount of CPU time taken. Note the benchmarks are ordered by the number of CNOTs in their circuit.}\label{tab:data}
	\\
	 \hline
	& \multicolumn{2}{|p{2.4cm}||}{\textsc{Original Circuit}} & \multicolumn{2}{p{2.4cm}||}{\textsc{Spectral}} & \multicolumn{2}{|p{2.2cm}||}{\textsc{CSU19 (Greedy)}}  &  \multicolumn{2}{p{2.2cm}||}{\textsc{CSU19 (Simple)}} & \multicolumn{2}{p{2.2cm}||}{\textsc{CSU19 (Extend)}} &  \multicolumn{2}{p{2.2cm}||}{\textsc{CSU19 (Qiskit)}}&  \multicolumn{2}{p{2.2cm}||}{\textsc{ZPW18 (A*)}}\\
	\textsc{Benchmark Circuit} & \textsc{Qubits} & \textsc{CNOTs} & \textsc{SWAP Count} & \textsc{CPU Time (sec)} & \textsc{SWAP Count} & \textsc{CPU Time (sec)} & \textsc{SWAP Count} & \textsc{CPU Time (sec)} & \textsc{SWAP Count} & \textsc{CPU Time (sec)} & \textsc{SWAP Count} & \textsc{CPU Time (sec)} & \textsc{SWAP Count} & \textsc{CPU Time (sec)} \\
\hline
\hline
\endfirsthead
	 \hline
& \multicolumn{2}{|p{2.4cm}||}{\textsc{Original Circuit}} & \multicolumn{2}{p{2.4cm}||}{\textsc{Lin (Spectral)}} & \multicolumn{2}{|p{2.2cm}||}{\textsc{CSU19 (Greedy)}}  &  \multicolumn{2}{p{2.2cm}||}{\textsc{CSU19 (Simple)}} & \multicolumn{2}{p{2.2cm}||}{\textsc{CSU19 (Extend)}} &  \multicolumn{2}{p{2.2cm}||}{\textsc{CSU19 (Qiskit)}}&  \multicolumn{2}{p{2.2cm}||}{\textsc{ZPW18 (A*)}}\\
\textsc{Benchmark Circuit} & \textsc{Qubits} & \textsc{CNOTs} & \textsc{SWAP Count} & \textsc{CPU Time (sec)} & \textsc{SWAP Count} & \textsc{CPU Time (sec)} & \textsc{SWAP Count} & \textsc{CPU Time (sec)} & \textsc{SWAP Count} & \textsc{CPU Time (sec)} & \textsc{SWAP Count} & \textsc{CPU Time (sec)} & \textsc{SWAP Count} & \textsc{CPU Time (sec)} \\
\hline
\hline
\endhead
 ex1\_226                      &                 6 &                5 &                   5   &                      0.3  &                   5   &                       0.9 &                   7   &                       1.6 &                   6   &                       1.5 &                   6.4 &                       0.9 &                     3 &                     0.002 \\
graycode6\_47                 &                 6 &                5 &                   0   &                      0.09 &                   4   &                       0.7 &                   6   &                       1   &                   5   &                       0.9 &                   4.2 &                       0.6 &                     1 &                     0.002 \\
xor5\_254                     &                 6 &                5 &                   5   &                      0.3  &                   5   &                       0.8 &                   7   &                       1.6 &                   6   &                       1.6 &                   6   &                       0.8 &                     3 &                     0.003 \\
ex-1\_166                     &                 3 &                9 &                   3   &                      0.1  &                   7   &                       1.5 &                   7   &                       2.2 &                   6   &                       2   &                   6.4 &                       1.2 &                     3 &                     0.004 \\
4gt11\_84                     &                 5 &                9 &                   3   &                      0.2  &                  11   &                       1.8 &                   8   &                       2.1 &                   6   &                       2   &                   6.4 &                       1.2 &                     5 &                     0.002 \\
\hline
\end{longtable}
	
\end{landscape}

\end{document}